\documentclass[aps,prd,twocolumn,showpacs,amsmath,nofootinbib]{revtex4-1}

 \usepackage{color}
\newcommand{\beqa}{\begin{eqnarray}}
\newcommand{\eeqa}{\end{eqnarray}}

\usepackage{graphicx}
\usepackage{epsfig}

\usepackage{graphicx}
\usepackage{dcolumn}
\usepackage{bm}%

\begin{document}

\title{Constraints on primordial black holes and curvature perturbations \\
from the global 21cm signal}

\author{Yupeng Yang$^{1,2,3}$}

\affiliation{$^1$School of Physics and Physical Engineering, Qufu Normal University, Qufu, Shandong, 273165, China\\
$^2$Collage of Physics and Electrical Engineering, Anyang Normal University, Anyang, Henan, 455000, China\\
$^3$Joint Center for Particle, Nuclear Physics and Cosmology, Nanjing, Jiangsu, 210093, China}

\begin{abstract}
The recent observations of the global 21cm 
signal by EDGES and gravitational waves by LIGO/VIGO have revived interest in PBHs. 
Motivated by these observations, many previous works focused on PBHs with lifetimes larger than the present age 
of the Universe. 
On one hand, taking into account the radiation from the gas accretion on to PBHs, 
the influences of the massive PBHs ($M_{\rm PBH}\sim \mathcal{O}(10^{2})M_{\odot}$) on the evolution of the intergalactic medium (IGM) has been investigated by previous works. 
On the other hand, considering the Hawking radiation, the similar effects of PBHs on the IGM have also been investigated by previous works for the less massive PBHs ($M_{\rm PBH}\gtrsim 10^{15}\rm g$). Different from previous works, 
we investigate the influence of PBHs on 
the evolution of the IGM for the mass range 
$6\times 10^{13} {\rm g} \lesssim M_{\rm PBH}\lesssim 3\times 10^{14} \rm g$. Since the lifetime of these PBHs 
is smaller than the present age of the Universe, they have evaporated by the present day.
Due to Hawking radiation, the heating effects of PBHs on the 
IGM can suppress the absorption amplitude of the global 21cm signal. 
In this work, by requiring that the differential brightness temperature of the global 21cm signals 
in the redshift range of $10\lesssim z \lesssim 30$, e.g., $\delta T_{b} \lesssim -100~\rm mK$, 
we obtain upper limits on the initial mass fraction of PBHs. We find that the strongest upper limit is 
$\beta_{\rm PBH} \sim 2\times 10^{-30}$. Since the formation of PBHs is related to primordial curvature perturbations, 
by using the constraints on the initial mass fraction of PBHs we obtain the upper limits on the 
power spectrum of primordial curvature perturbations for the scale range $8.0\times 10^{15}\lesssim k \lesssim 1.8\times 10^{16}~\rm Mpc^{-1}$, corresponding to the mass range considered here. 
We find that the strongest upper limit is $\mathcal P_{\mathcal R}(k) \sim 0.0046$. 
By comparing with previous works, we find that for the mass range (or the scale range)
investigated in this work the global 21cm signals or the 21cm power spectrum should give the strongest upper limits 
on the initial mass fraction of PBHs and on the power spectrum of primordial curvature perturbations. 

\end{abstract}

\maketitle

\section{introduction} 
Primordial black holes (PBHs) can form in the early epoch of 
the Universe if there are large density perturbations. 
Depending on their mass, PBHs can emit different particles 
via Hawking radiation~\cite{carr,pbhs_emit_3,pbhs_emit_2,pbhs_emit_1}, 
which then interact with other particles in the Universe. 
The evolution of the intergalactic medium (IGM) is changed due to these interactions, 
and these changes can influence astrophysical observations, e.g. the global 21cm signal~\cite{mack_21,yinzhema}. 

The Experiment to Detect the Global Epoch of Reionization Signature (EDGES) 
has reported the observation of the global 21cm signal, which shows an absorption feature 
with an amplitude of $T_{\rm 21}\sim 500\ \rm mK$ 
centered at redshift $z\sim 17$ and is about a factor of 2 larger than expected~\cite{edges-nature}. 
According to the theory, the global 21cm signal is controlled 
by the evolution of the kinetic temperature ($T_{k}$), 
the CMB thermodynamic temperature ($T_{\rm CMB}$) and 
the spin temperature ($T_{s}$). One way to explain the large amplitude observed by the EDGES experiment 
is to require the IGM to be cooler than expected, which could be caused 
by, e.g., the interactions between dark matter particles and baryons~\cite{Barkana:2018lgd}. 
Another way is to enhance the intensity of the radio background at low frequencies, 
which can be satisfied by possible radio sources~\cite{Feng:2018rje,prd-edges}. 
In general, any additional source, e.g. the dark matter annihilation or decay, 
will heat the IGM and increase the kinetic temperature~\cite{prd-edges,DM_2015,xlc_decay,lz_decay,chi_1,chi_2,energy_function,
yinzhema}. 
In order to be consistent with the observational results of the EDGES experiment, the properties of 
the dark matter particles should be constrained
~\cite{yinzhema,DAmico:2018sxd,prd-edges,Kovetz:2018zes,Bhatt:2019qbq,Berlin:2018sjs,Barkana:2018cct}. 
As mentioned above, due to Hawking radiation the evolution of the IGM can be influenced 
by PBHs, and therefore the mass fraction of PBHs 
can be constrained by the global 21cm signal~\cite{mack_21,yinzhema}. 
In Ref.~\cite{yinzhema}, the authors focused on PBHs in the mass range $M_{\rm PBH} \gtrsim 10^{15} \rm g$ and investigated their influence 
on the evolution of the IGM. The lifetime of a PBH with a mass $M_{\rm PBH} \gtrsim 10^{15}\rm g$ is longer than 
the age of the Universe, and therefore these PBHs have not evaporated by the present day. 
Taking into account the global 21cm signal, the authors of Ref.~\cite{yinzhema} found the upper limits on the present mass fraction 
of PBHs depending on the masses of PBHs, e.g., $f_{\rm PBH} \sim 10^{-9}$ 
for $M_{\rm PBH} \sim 10^{15} \rm g$. Different from Ref.~\cite{yinzhema}, here 
we focus on PBHs in the mass range $10^{13} {\rm g}\lesssim  
M_{\rm PBH} \lesssim 10^{14}\rm g$, which have evaporated 
in the redshift range $6 \lesssim z \lesssim 1100$. 
In Ref.~\cite{mnras}, the authors investigated the influence of PBHs on the evolution of the IGM for a similar mass range 
and obtained the upper limits on the initial mass fraction of PBHs 
using the Planck-2015 data, e.g., $\beta_{\rm PBH} \sim 10^{-28}$ 
for $M_{\rm PBH} \sim 10^{14} \rm g$. For other methods and more detailed discussions on the constraints 
of initial mass fraction of PBHs see, e.g., Ref.~\cite{carr} and references therein.

PBHs can be used to investigate the relevant issues of the early Universe. For example, the initial mass fraction of PBHs 
is related to primordial curvature perturbations~\cite{Josan:2009,Sato-Polito:2019hws}. 
A nearly scale-invariant spectrum of primordial curvature perturbations has been predicted 
by many inflation models~\cite{model_1}. 
The most robust constraints on power spectrum of the primordial curvature perturbations, $\mathcal{P}_\mathcal{R}(k)$, 
are from the observations and studies of CMB, Lyman-$\alpha$ forest, and large-scale structure~\cite{cmb_2,lyman,large}, 
and these constraints apply on scales $10^{-4}\ 
\lesssim k \lesssim 1\ \mathrm{Mpc^{-1}}$ with a nearly invariant value of 
$\mathcal{P}_\mathcal{R}(k) \sim 10^{-9}$. Since PBHs originate from 
the collapse of early density perturbations 
they can be used to constrain primordial curvature perturbations. The upper limits on 
$\mathcal{P}_\mathcal{R}(k)$ from research on PBHs apply on for scales $k \lesssim 10^{20}\ \mathrm{Mpc^{-1}}$ and 
are about $\sim 7$ orders of magnitude weaker than that from the CMB, Lyman-$\alpha$ forest, and large-scale structure~\cite{Josan:2009,pbhs_2016,Dalianis:2018ymb}. For scales in the range $5 \lesssim k \lesssim 10^{8}\ \rm Mpc^{-1}$, 
the upper limits on $\mathcal{P}_\mathcal{R}(k)$ can be obtained from the studies on the ultracompact dark matter minihalos, and these limits are about $\sim 3$ orders of magnitude stronger than that from PBHs~\cite{Josan,Bringmann_1,fangdali,scott_2015,yyp_neutrino,epl,Nakama_2018}. 
In Ref.~\cite{prl_2}, taking into account the Silk damping effects in the early Universe, the authors found 
an upper limit of $\mathcal{P}_\mathcal{R}(k)\sim 0.06$ for the 
scale range $10^{4}\ \lesssim k \lesssim 10^{5}\ \rm Mpc^{-1}$. 
Utilizing the Planck-2015 data, the authors of~\cite{mnras} obtained upper limits on $\mathcal{P}_\mathcal{R}(k)$ 
for the scale range $8.9 \times 10^{15} \lesssim k \lesssim 2.8\times 10^{16}~\rm  Mpc^{-1}$. In this work, 
using the upper limits on the initial mass fraction of PBHs obtained from 
the global 21cm signals, we obtain upper limits on $\mathcal{P}_\mathcal{R}(k)$ 
for the scale range $8.0\times 10^{15} \lesssim k \lesssim 1.8\times 10^{16}~\rm Mpc^{-1}$. 

This paper is organized as follows. In Sec. II we discuss the basic properties of PBHs 
and their effects on the evolution of the IGM due to Hawking radiation. 
The influence of PBHs on the global 21cm signal are investigated in Sec. III. 
We obtain upper limits on the initial mass fraction of PBHs and the power spectrum of the primordial curvature 
perturbations in Sec. IV. The conclusions and discussions are given in Sec. V.

\section{the influence of PBHs on the evolution of the IGM}

\subsection{The basic properties of PBHs}

In this section we briefly review the basic properties of PBHs. For more detailed discussions, one can refer to, 
e.g., Refs.~\cite{pbhs_emit_3,pbhs_emit_2,pbhs_emit_1,pbhs_2016,pbhs_review,carr,Carr_2020,carr2020constraints} and references therein. 

In the early epoch of the Universe, PBHs can form if there are large density perturbations. 
The needed amplitude of large density perturbation is generally of~$\delta \rho/\rho \gtrsim 0.3$~\cite{Green:1997sz}. 
According to the theory, a PBH can radiate thermally 
with temperature~\cite{pbhs_emit_3,pbhs_emit_2,pbhs_emit_1,Josan:2009,carr} 

\beqa
T_{\rm PBH}=\frac{1}{ 8\pi GM_{\rm PBH}}\approx \left(
\frac{M_{\rm PBH}}{10^{13}\rm g}\right)^{-1} \rm GeV.
\label{eq:tem_mass}
\eeqa 

The mass of a PBH changes with time due to Hawking radiation. The mass loss rate of a black hole can be expressed 
as~\cite{Josan:2009} 

\beqa
\frac{{\rm d}M_{\rm BH}}{{\rm d}t} = -5.34\times 10^{25}
f\left(M_{\rm BH}\right)\left(\frac{M_{\rm BH}}{\rm g}\right)^{-2} \rm g\ s^{-1},
\label{eq:mass_loss}
\eeqa
where $f(M_{\rm BH})$ measures the number of particle species that are 
emitted directly. $f(M_{\rm BH})$ can be calculated exactly~\cite{carr} and 
in this work we use the fitted formula used in, e.g., Ref.~\cite{Tashiro:2008sf},

\beqa
f(M_{\rm BH})=&&1.569 +0.569 e^{\frac{-0.0234}{T_{\rm BH}}} + 3.414e^{\frac{-0.066}{T_{\rm BH}}} \nonumber \\
	     && + 1.707e^{\frac{-0.11}{T_{\rm BH}}} + 0.569e^{\frac{-0.394}{T_{\rm BH}}} \nonumber \\
	     && + 1.707e^{\frac{-0.413}{T_{\rm BH}}}  + 1.707e^{\frac{-1.17}{T_{\rm BH}}} \nonumber \\
	     && + 1.707e^{\frac{-22}{T_{\rm BH}}} + 0.963e^{\frac{-0.1}{T_{\rm BH}}}
\eeqa
where $T_{\rm BH}$ is determined by Eq.~(\ref{eq:tem_mass}). For the mass range considered here, $f(M_{\rm PBH})$ 
is in the range $2.5 \lesssim f(M_{\rm BH}) \lesssim 6.1$, corresponding to the temperature range $33 \rm ~MeV \lesssim T_{\rm PBH} \lesssim 167\rm ~MeV$. 
For this range, pions, muons, quarks (up, down, and strange), and gluons will be emitted~\cite{pbhs_2016,pbh_quaks}, 
and hadrons will be produced after the emission of quarks and gluons through the process of fragmentation. 
In this work, we will investigate the influence of PBHs on the evolution of the IGM. Based on previews works, it has been found that 
the main influence of PBHs on the IGM is due to electrons, positrons, and photons~\cite{mack_21,xlc_decay}. 
Therefore, following previous works, we consider electrons, positron, and photons that are emitted directly by PBHs or 
produced indirectly through the decay of, e.g., muons, pions, and other hadrons~\cite{mack_21,yinzhema,pbh_quaks,carr}.

The lifetime of a 
PBH with a fixed mass, $\tau_{\rm PBH}$, can be obtained by integrating 
Eq.~(\ref{eq:mass_loss}). One good approximation of the lifetime can 
be written as~\cite{Josan:2009} 

\beqa
\tau_{\rm PBH} \approx 3\times 10^{14} \left(\frac{M_{\rm PBH}}
{10^{14}\rm g}\right)^{3}f(M_{\rm PBH})^{-1}~s.
\label{eq:tau}
\eeqa

According to Eq.~(\ref{eq:tau}), the lifetime of a PBH with 
mass of $M_{\rm PBH}\sim 5\times 10^{14}~\rm g$ is equal to the age of the 
Universe, $t \sim 13.7~ \rm Gyr$~\cite{planck-2018}. Therefore, for masses of 
$M_{\rm PBH} < 5\times 10^{14}\rm g$, PBHs have evaporated 
by the present day. The final stages of PBHs, 
e.g., a stable Planck mass relic or a connection to extra dimensions, 
have been discussed in previous works~\cite{Kavic:2008qb,Bowick:1988xh,Coleman:1991sj}. 
Following Ref.~\cite{mack_21}, in this work we assume that the Hawking evaporation stops at the final stages. 


\subsection{The evolution of the IGM including PBHs}

There are interactions between the particles emitted from PBHs and that existed in the Universe. 
Due to these interactions the evolution of the IGM is changed. The main influence 
of the interactions on the IGM are heating, ionization and excitation~\cite{lz_decay,xlc_decay,yinzhema,mnras,DM_2015,prd-edges,Belotsky:2014twa}. 
For our purposes, the ionization fraction ($x_e$) and the temperature of the IGM ($T_{k}$) 
are mainly used to study the evolution of the IGM. The evolutions of $x_e$ and $T_k$ with the redshift 
can be written as~\cite{mnras,DM_2015,xlc_decay,lz_decay} 

\beqa
(1+z)\frac{dx_{e}}{dz}=\frac{1}{H(z)}\left[R_{s}(z)-I_{s}(z)-I_{\rm add}(z)\right],
\eeqa

\beqa
(1+z)\frac{dT_{k}}{dz}=&&\frac{8\sigma_{T}a_{R}T^{4}_{\rm CMB}}{3m_{e}cH(z)}\frac{x_{e}}{1+f_{\rm He}+x_{e}}
(T_{k}-T_{\rm CMB})\\ \nonumber
&&-\frac{2}{3k_{B}H(z)}\frac{K_{\rm add}}{1+f_{\rm He}+x_{e}}+T_{k}, 
\eeqa
where $R_{s}(z)$ is the recombination rate, $I_{s}(z)$ is the ionization rate 
caused by the standard sources. 
$I_{\rm add}$ and $K_{\rm add}$ are the ionization rate and heating rate caused by the additional sources. 
For our purposes, $I_{\rm add}$ and $K_{\rm add}$ are caused by PBHs, i.e. $I_{\rm add}\equiv I_{\rm PBH}$ and 
$K_{\rm add}\equiv K_{\rm PBH}$, and they can be written as~\cite{lz_decay,xlc_decay,mnras,yinzhema,DM_2015} 

\beqa
I_{\rm PBH} = \chi_{i} f\frac{1}{n_b}\frac{1}{E_0}\times
\frac{{\rm d}E}{{\rm d}V{\rm d}t}\bigg|_{\rm PBH}
\label{eq:I}
\eeqa
\beqa
K_{\rm PBH} = \chi_{h} f\frac{1}{n_b}\times \frac{{\rm d}E}{{\rm d}V{\rm d}t}\bigg|_{\rm PBH} 
\label{eq:K}
\eeqa
 
The energy injection rate per unit volume due to PBHs can be written as

\beqa
\frac{{\rm d}E}{{\rm d}V{\rm d}t}\bigg|_{\rm PBH} = \frac{1}{M_{\rm PBH}}\frac{{\rm d}M_{\rm PBH}}{{\rm d}t}
n_{\rm PBH}(z),
\eeqa
where $n_{\rm PBH}(z)$ is the number density of PBHs at redshift $z$. 
The initial mass fraction of PBHs, $\beta_{\rm PBH}$, can be written as~\cite{Josan:2009} 

\beqa
\beta_{\rm PBH} \equiv \frac{\rho_{\rm PBH}^{i}}{\rho_{\rm crit}^{i}}=
\frac{\rho_{\rm PBH}^{\rm eq}}{\rho_{\rm crit}^{\rm eq}}\left(\frac{a_i}{a_{\rm eq}}\right),
\eeqa
where $a=1/(1+z)$ is the scale factor, $\rho_{\rm PBH}^{i}$ ($\rho_{\rm PBH}^{\rm eq}$) and 
$\rho_{\rm crit}^{i}$ 
($\rho_{\rm crit}^{\rm eq}$) are the mass density of PBHs
and the critical density of the Universe at the time of PBH formation 
(matter-radiation equality), respectively. 
The scale factor is related to the horizon mass $M_{\rm H}$ as~\cite{Josan:2009}

\beqa
\frac{a_i}{a_{\rm eq}} = \left(\frac{g_{\star}^{\rm eq}}{g_{\star}^{i}}\right)^{1/12}
\left(\frac{M_{\rm H}}{M_{\rm H}^{\rm eq}}\right)^{1/2},
\eeqa
where $g_{\star}^{\rm eq}\approx 3$ and $g_{\star}^{i}\approx 100$ are 
the total number of effectively massless degrees of freedom at 
the epoch of matter-radiation equality and PBH formation, respectively. 
With the above equations, the number density of PBHs can be rewritten as
\footnote{Here we assume a monochromatic mass fraction for PBHs, and we will investigate
 the extended mass spectrum for PBHs in detail in future work.} 

\beqa
n_{\rm PBH}(z) &&= \beta_{\rm PBH}\left(\frac{1+z}{1+z_{\rm eq}}
\right)^{3}\frac{\rho_{\rm crit}^{\rm eq}}{M_{\rm PBH}^{i}}
\left(\frac{g_{\star}^{i}}{g_{\star}^{\rm eq}}\right)^{\frac{1}{12}}
\left(\frac{M_{\rm H}^{\rm eq}}{M_{\rm H}}\right)^{\frac{1}{2}}\nonumber  \\
&&=1.46\times 10^{-4}\beta_{\rm PBH}\left(1+z\right)^3 \left(\frac{M_{\rm PBH}^i}{\rm g}\right)^{-3/2}
\eeqa
where $M_{\rm PBH}^{i}$ is the mass of PBH at the formation time and 
we have used the relations and values, 
$M_{\rm H}^{\rm eq} = 1.3\times 10^{49}\left(\Omega_{\rm m}h^{2}\right)^{-2}~\rm g$, 
$M_{\rm PBH}^{i}=f_{\rm M}M_{\rm H}$ with $f_{\rm M}=(1/3)^{3/2}$, 
$\rho_{\rm crit}=1.88\times 10^{-29}h^{2}~\rm g ~cm^{-3}$ and $z_{\rm eq} = 3403$~\cite{Josan:2009,planck-2018}. 

\begin{figure}
\epsfig{file=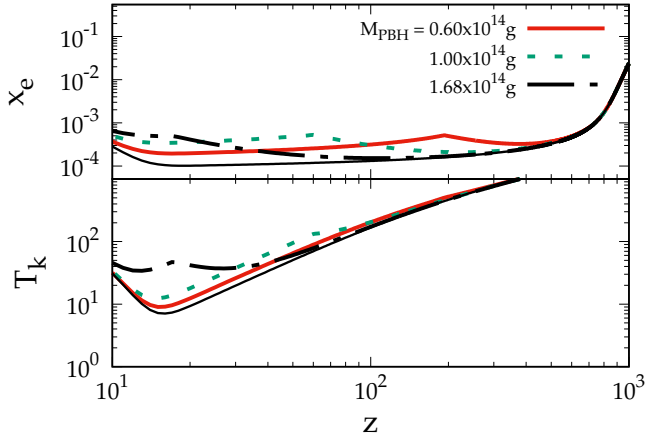,width=0.5\textwidth}
\caption{The evolutions of $x_e$ and $T_k$ for different PBH masses: $M_{\rm PBH}=0.60\times 10^{14}\rm g$ (red solid line), 
$1.00\times 10^{14}\rm g$ (green dotted line) and $1.68\times 10^{14}\rm g$ (black dot-dashed line). Here we have set the initial 
mass fraction of PBHs as $\beta_{\rm PBH} = 10^{-29}$. For comparison, we also plot the evolutions of $x_e$ and $T_k$ for the case with no PBHs(thin black solid line).}
\label{fig:xe_tk}
\end{figure}

In Eqs.~(\ref{eq:I}) and~(\ref{eq:K}), the factor $f$ stands for the energy fraction 
injected into the IGM due to Hawking radiation and it is a function of 
redshift~\cite{binyue,Cumberbatch:2008rh,energy_function,Slatyer:2015kla,chi_1}. 
In general, the energy that can be injected into the IGM is mainly caused by electrons and photons~\cite{xlc_decay,mack_21}. $\chi_{i(h)}$ are the fractions of the energy deposited into the IGM for ionization (heating)~\cite{xlc_decay,mack_21,lz_decay,DM_2015,Slatyer:2015kla}. According to theories on structure formation, the first stars could be formed after the redshift of $z\sim 30$. These stars became the main sources for ionization, 
and in this work we adopt the model used 
in Refs.~\cite{prd-edges,binyue}. We have modified the public code RECFAST in CAMB\footnote{https://camb.info/} 
to include the influence of PBHs. The evolutions of $x_e$ and $T_k$ 
are shown in Fig.~\ref{fig:xe_tk}. For this figure, we have set the initial mass fraction of PBHs as $\beta_{\rm PBH} = 10^{-29}$. 
Due to the influence of PBH, in general, $x_e$ and $T_k$ are larger than that for the case with no PBH, which can be seen clearly in 
Fig.~\ref{fig:xe_tk} especially after 
the redshift $z\sim 600$. The details of the evolutions of $x_e$ and $T_k$ are different depending on the mass of PBHs. 
Since we have assumed that PBHs stop 
evaporating at the final stage, therefore, there are inflections in Fig.~\ref{fig:xe_tk}. 
For the mass range $0.6\times 10^{14}{\rm ~g}\lesssim M_{\rm PBH}\lesssim 10^{14}{\rm ~g}$, PBHs evaporate in the redshift range 
$30\lesssim z\lesssim 300$. The $x_e$ and $T_k$ reach their largest values at redshift of $z^{\prime}$ (inflections in Fig.~\ref{fig:xe_tk}) 
corresponding to 
the lifetime of PBHs. After the redshift $z^{\prime}$, the evolutions of $x_e$ and $T_k$ tend to follow the cases without PBHs. 
Similar evolutions can be found for the mass range of $10^{14}{\rm ~g} \lesssim M_{\rm PBH}\lesssim 3\times 10^{14} \rm ~g$, 
in which PBHs evaporate in the redshift range $6\lesssim z\lesssim 30$.

\section{The influence of PBHs on the global 21cm signal}

In the early epoch, the Universe is in the ionized phase and the temperature of the IGM is very high. With the expansion 
of the Universe, the temperature decreases and the hydrogen atoms form due to the combination of protons and 
electrons at redshift $z\sim 1100$. The 21cm line is related to the transition between the triplet and singlet levels 
of the ground state of the hydrogen atom. The transition energy between the two levels is $E = 5.9\times 10^{-6} ~\rm eV$, corresponding to the wavelength of photon $\lambda = 21 ~\rm cm$. The spin 
temperature, $T_s$, which is used to describe the transition, is defined as~\cite{Pritchard:2011xb,Furlanetto:2006jb} 

\beqa
\frac{n_1}{n_0}=3\mathrm{exp}\left(-\frac{T_{\star}}{T_s}\right),
\eeqa
where $n_0$ and $n_1$ are the number densities of hydrogen atoms in triplet and singlet states, $T_{\star}=0.068~\rm K$ is the temperature corresponding to the transition energy. The spin temperature $T_s$ is mainly effected by 
(i) background photons; (ii) collisions between the hydrogen atoms and other particles; (iii) resonant scattering 
of $\rm Ly\alpha$ photons named Wouthuysen-Field effect~\cite{Pritchard:2011xb,Furlanetto:2006jb}. Considering the cosmic microwave background as the main part 
of the background photons, the spin temperature can be written as~\cite{binyue,Cumberbatch:2008rh}

\beqa
T_{s} = \frac{T_{\rm CMB}+(y_{\alpha}+y_{c})T_{k}}{1+y_{\alpha}+y_{c}},
\eeqa
where $y_{\alpha}$ corresponds to the Wouthuysen-Field effect and we adopt the form used 
in e.g. Refs.~\cite{binyue,mnras,Kuhlen:2005cm}, 

\beqa
y_{\alpha} = \frac{P_{10}T_{\star}}{A_{10}T_{k}}e^{-0.3\times(1+z)^{0.5}T_{k}^{-2/3}\left(1+\frac{0.4}{T_{k}}\right)^{-1}},
\eeqa
where $A_{10}=2.85\times 10^{-15}s^{-1}$ is the Einstein coefficient of the hyperfine spontaneous transition. 
$P_{10}$ is the de-excitation rate of the hyperfine triplet state due to $\rm Ly{\alpha}$ scattering~\cite{Pritchard:2011xb,Furlanetto:2006jb}. $y_c$ corresponds to 
the collisions between hydrogen atoms and other particles~\cite{binyue,prd-edges,Kuhlen:2005cm,Liszt:2001kh,epjplus-2}, 

\beqa
y_{c} = \frac{(C_{\rm HH}+C_{\rm eH}+C_{\rm pH})T_{\star}}{A_{10}T_{k}},
\eeqa  
where $C_{\rm HH, eH, pH}$ are the de-excitation rate and we adopt the forms used in Refs.~\cite{prd-edges,epjplus-2,Kuhlen:2005cm,Liszt:2001kh}. 

In general, the differential brightness temperature, $\delta T_b$, is used to describe the global 21cm signals, which can be written 
as~\cite{Cumberbatch:2008rh,Ciardi:2003hg,prd-edges} 

\beqa
\delta T_b = ~&& 26(1-x_e)\left(\frac{\Omega_{b}h}{0.02}\right)\left[\frac{1+z}{10}\frac{0.3}{\Omega_{m}}\right]^{\frac{1}{2}}\\ \nonumber
&&\times \left(1-\frac{T_{\rm CMB}}{T_s}\right)~\rm mK.
\eeqa

\begin{figure}
\epsfig{file=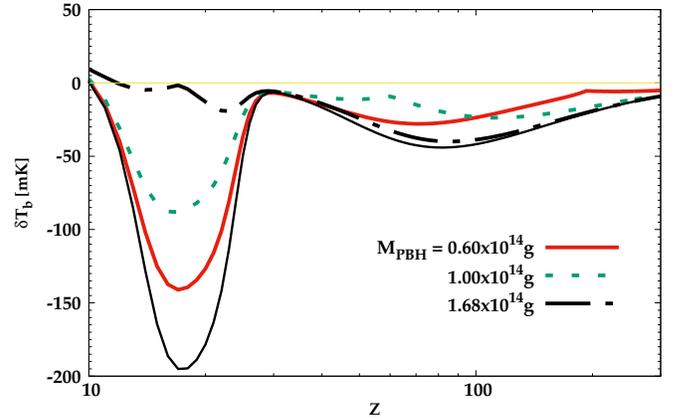,width=0.5\textwidth}
\caption{The evolutions of the differential brightness temperature $\delta T_b$ in the redshift range $10\lesssim z\lesssim 300$ for different PBH masses, $M_{\rm PBH}=0.60\times 10^{14}\rm g$ (red solid line), 
$1.00\times 10^{14}\rm g$ (green dotted line) and $1.68\times 10^{14}\rm g$ (black dot-dashed line). Here we have set the initial 
mass fraction of PBHs as $\beta_{\rm PBH} = 10^{-29}$. For comparison, we also plot the evolution of $\delta T_b$ for the case with no PBHs(thin black solid line).}
\label{fig:delta_eg}
\end{figure}

Using the above equations, in Fig.~\ref{fig:delta_eg} we display the evolutions of $\delta T_b$ in the redshift range $10\lesssim z\lesssim 300$ for different PBH masses. For comparison, 
the case with no PBHs is also shown (thin black solid line). From Fig.~\ref{fig:delta_eg}, it can be seen that due to the influence of PBHs the absorption amplitude of the global 21cm signal is suppressed, which is caused mainly by the heating effects of PBHs on the IGM. Similar results can also be found in e.g. Refs.~\cite{mack_21,binyue,prd-edges,yinzhema}. Depending on the different PBH masses, the global 21cm signal exhibits different features and there are also inflections in the plots. 
For lower mass, e.g. $M_{\rm PBH} = 0.6\times 10^{14} \rm ~g$ (red solid line), PBHs evaporate at the redshift $z\sim 200$. 
For the redshifts $z\lesssim 200$, 
the evolutions of $x_e$ and $T_k$ (Fig.~\ref{fig:xe_tk}) tend to follow the case with no PBHs but are still larger than that case (thin black solid line). As a result, the amplitude of $\delta T_b$ becomes 
smaller than that for the case with no PBHs. For larger mass , e.g. $M_{\rm PBH} = 1.68\times 10^{14} \rm g$ (black dot-dashed line), PBHs evaporate in the redshift range $10\lesssim z\lesssim 30$. 
For this case, the absorption amplitude of the global 21cm signal is strongly suppressed, and the absorption trough tends to become an emission peak. 
As shown in Fig.~\ref{fig:delta_eg}, there are two absorption features in the global 21cm signal. 
One is in the redshift range $30 \lesssim z\lesssim 300$ and another one appears in the redshift range $10 \lesssim z\lesssim 30$. For the mass range considered by us, 
PBHs have effects on the global 21cm signals in both redshift ranges. Specifically, the main effects are in the higher (lower) redshift range for the smaller (larger) PBHs 
depending on their lifetime. For our purposes, inspired by the observational results of the EDGES experiment, we have focused on the redshift range $10\lesssim z \lesssim 30$. 
As shown in Fig.~\ref{fig:delta_eg}, for this redshift range, larger PBHs have stronger effects on the global 21cm signal compared with that in the redshift range $30\lesssim z\lesssim 300$. 
For the plots, we have set the initial mass fraction of PBHs as $\beta_{\rm PBH} = 10^{-29}$, and PBHs with larger initial mass fractions (fixed mass) should have stronger effects on the global 21cm signal. 
Therefore, in view of the observational results of the EDGES experiment, the initial mass fraction of PBHs should have upper bounds, and this issue will be discussed in the following section.

\section{Constraints on PBHs and curvature perturbations}
\subsection{Constraints on the initial mass fraction of PBHs}

\begin{figure}
\epsfig{file=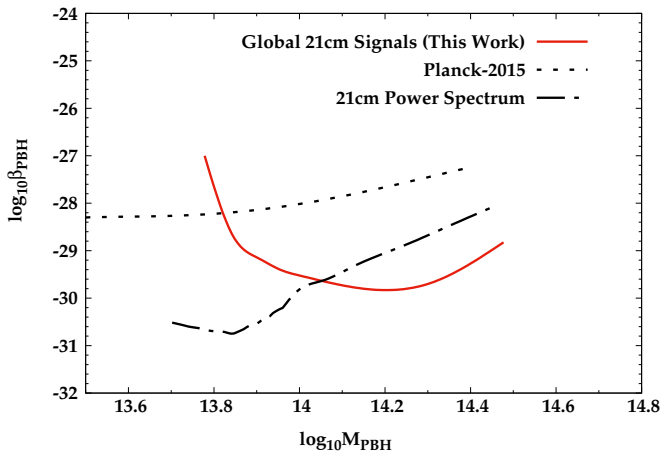,width=0.5\textwidth}
\caption{Constraints on the initial mass fraction of PBHs, $\beta_{\rm PBH}$, 
by requiring the differential brightness temperature of the global 21cm signal to be 
in the redshift range $10\lesssim z\lesssim 30$ as $\delta T_{b}\lesssim -100 \rm mK$ (red solid line). For comparison, 
the upper limits from the Placnk-2015 data~\cite{mnras} (black dotted line) 
and the potential 21cm power spectrum~\cite{mack_21} (black dashed-dotted line) are also shown.}
\label{fig:dPBHSLCF}
\end{figure}

As the discussed above, PBHs with different masses have a different significant influence on the global 21cm signal. 
As shown in Fig.~\ref{fig:delta_eg}, the main influence is a suppression of the amplitude of the absorption trough. 
Moreover, the absorption trough could disappear or become an emission peak due to the effects of PBHs with large initial mass fractions. 
Therefore, the global 21cm signal can be used to investigate the abundance of PBHs. 
Recently, the global 21cm signal with a large absorption trough was observed by the EDGES experiment 
at the redshift $z\sim 17$. 
Following previous works~\cite{yinzhema,DAmico:2018sxd}, we obtain the upper limits on the initial mass fraction of PBHs, $\beta_{\rm PBH}$, by requiring the differential brightness temperature of the global 21cm signals to be 
$\delta T_{b} \lesssim -100~\rm mK$. In Fig.~\ref{fig:dPBHSLCF}, we display the upper limits on $\beta_{\rm PBH}$ for the mass 
range $6\times 10^{13} {\rm g} \lesssim M_{\rm PBH}\lesssim 3\times 10^{14} \rm g$ (red solid line). From this plot, it can be seen that the strongest upper limit on the initial mass fraction of PBHs is $\beta_{\rm PBH} \sim 2\times 10^{-30}$. 
Because we have considered the effects of PBHs on the global 21cm signal in the redshift range 
$10\lesssim z \lesssim 30$, the strongest upper limit appears for larger masses (longer lifetime) in the mass range considered here. 

In Ref.~\cite{mack_21}, using the expected 21cm power spectrum observed by SKA, the authors obtained the potential upper limits on $\beta_{\rm PBH}$ and the strongest upper limit is $\beta_{\rm PBH} \sim 2\times 10^{31}$, 
which is also displayed in 
Fig.~\ref{fig:dPBHSLCF} (black dashed-dotted line). Different from this work, the authors 
of Ref.~\cite{mack_21} focused on the influence 
of PBHs on the 21cm signal in the redshift range $30\lesssim z \lesssim 300$. Therefore, 
as shown in Fig.~\ref{fig:dPBHSLCF}, for the mass range considered here the constraints 
on $\beta_{\rm PBH}$ are stronger for smaller masses (shorter lifetime) than those for larger masses (longer lifetime). 

The influence of PBHs on the IGM can also effect the anisotropy of the CMB. Therefore, 
CMB observations can also be used to 
investigate the initial mass fraction of PBHs. 
Utilizing the Planck-2015 data, the authors of Ref.~\cite{mnras} obtained the upper limits on $\beta_{\rm PBH}$ for the mass range $2.8\times 10^{13} {\rm g} \lesssim M_{\rm PBH}\lesssim 2.5\times 10^{14} \rm g$, and they found that the strongest limit 
is $\beta_{\rm PBH} \sim 4\times 10^{-29}$ (black dotted line in Fig.~\ref{fig:dPBHSLCF}). Since the constraints 
on $\beta_{\rm PBH}$ from the CMB data 
are mainly from the high redshift range, the strongest upper limit on $\beta_{\rm PBH}$ appears 
for smaller masses\footnote{PBHs with smaller masses evaporate before recombination ($z\sim 1000$). Therefore, as shown in Fig.~\ref{fig:dPBHSLCF}, there is a cut on the masses of PBHs.}, which can be seen in Fig.~\ref{fig:dPBHSLCF} and 
is similar to that of the 21cm power spectrum.

PBHs with masses $M_{\rm PBH} \lesssim 6\times 10^{13}\rm g$ evaporate before the recombination. Therefore, 
for this mass range the constraints on the initial mass fraction of PBHs are mainly from the BBN and CMB distortions~\cite{Tashiro:2008sf,carr}, 
and the strongest upper limit is about $\beta_{\rm PBH} \sim 10^{-24}$. For large masses 
$M_{\rm PBH} \gtrsim 6\times 10^{15}~{\rm g}$, the constraints on $\beta_{\rm PBH}$ are mainly from the lensing effect~\cite{carr}. 
The radiation from the accretion onto PBHs can also affect the evolution of the IGM, and the CMB data can also be used 
to obtain the upper limits on $\beta_{\rm PBH}$~\cite{Ricotti:2007au,Chen:2016pud,Ali-Haimoud:2016mbv,Poulin:2017bwe}. The recent observations of gravitational waves provide an important way 
to constrain the initial mass fraction of PBHs for $M_{\rm PBH}\sim \mathcal{O}(10^{2})M_{\odot}$~\cite{Wang:2016ana,Bird:2016dcv}. 
For other methods and more detailed 
discussions about the constraints on the initial mass fraction of PBHs, one can refer to 
e.g. Refs.~\cite{carr,Josan:2009,He:2002vz,Green:1997pr,Hektor:2018qqw,Murgia:2019duy,Petkov:2019edm,Chluba:2019nxa,
Poulter:2019ooo,Mena:2019nhm,Wang:2018ydd,Nakama:2017xvq,Inoue:2017csr,Gaggero:2016dpq,Lu:2019ktw,cai2020constraints} and references therein.

As discussed above, the upper limits of the initial mass fraction of PBHs are different for the different astrophysical observations. 
Since the main influence of PBHs on the CMB are on the higher redshifts, the upper limit on the initial mass fraction of PBH with lower mass (shorter lifetime) 
is stronger than that of PBHs with larger mass. Compared with the smaller mass, as shown in Fig.~\ref{fig:delta_eg}, 
PBHs with larger mass have a significant influence 
on the global 21cm signal observed by the EDGES experiment at the redshift $z\sim 17$. 
Therefore, different from the constraints from the CMB, the upper limit on the initial mass fraction of 
PBHs with larger mass (longer lifetime) is stronger than that of PBHs with smaller mass. In particular, 
PBHs with mass $M_{\rm PBH} \lesssim 10^{13.8}\rm g$ evaporate at redshift $z > 30$. Therefore, the observational results of the EDGES experiment cannot give stringent upper limits on the initial mass fraction of PBHs. 
On the other hand, since the lifetime of PBHs with mass $M_{\rm PBH} \gtrsim 10^{14.4}\rm g$ 
is larger than the present age of the Universe, 
the observational results of the EDGES experiment also cannot give 
stringent upper limits on the initial mass fraction of PBHs. 
From Fig.~\ref{fig:dPBHSLCF}, it can be seen  that the strongest upper limits correspond to the intermediate mass range, in which PBHs 
evaporate in the redshift range $10\lesssim z\lesssim 30$. 

Moreover, for the mass range $6\times 10^{13} {\rm g} \lesssim M_{\rm PBH}\lesssim 3\times 10^{14} \rm g$, 
comparing with other constraints, the strongest upper limit on the initial mass fraction of 
PBHs comes from the observations of the global 21cm signal or the future 21cm power spectrum. In this mass range,  constraints can also be obtained through studies of the extragalactic photon background, extragalactic antiprotons and neutrinos~\cite{carr,pbhs_2016}, but the constraints are weaker than the 21cm constraints~\cite{mack_21,carr}.

\subsection{Constraints on primordial curvature perturbations}

PBHs can form via the collapse of the large density perturbations present in the early epoch of the Universe. 
The density perturbations can be Gaussion or non-Gaussion~\cite{forma_pbhs_2}. In this work, we have considered 
Gaussion perturbations, 
and in this case, in light of the Press-Schechter theory~\cite{ps}, the initial mass fraction $\beta_{\rm PBH}$ can be written as~\cite{Josan:2009}
\beqa
\beta_{\rm PBH} = \frac{2M_{\rm PBH}}{M_{\rm H}}\int^{1}_{\delta_{c}}p(\delta_{H}(R))d\delta_{H}(R)
\eeqa
where $M_{\rm PBH} = f_{\rm M}M_{\rm H}$, $M_{\rm H}$ is the horizon mass at the formation time of PBHs, $f_{\rm M}$ is the fraction of 
the horizon mass that collapses into PBHs. 
$\delta_{c}=\delta \rho/\rho$ is the critical value of the density perturbation that can form PBHs and here we set 
$\delta_{c} = 1/3$~\cite{pbhs_review}. 
$\delta_{H}(R)$ is the smoothed density contrast at horizon crossing with $R=(aH)^{-1}$. 
$p(\delta_{H}(R))$ is the probability distribution of the smoothed density contrast. For the Gaussian perturbations, $p(\delta_{H}(R))$ 
can be written as 

\beqa
p(\delta_{c}(R))=\frac{1}{\sqrt{2\pi}\delta_{H}(R)}{\rm exp}\left(-\frac{\delta_{H}^{2}(R)}{2\sigma^{2}_{H}(R)}\right), 
\eeqa
where $\sigma_{H}(R)$ is the mass variance in the form of 

\beqa
\sigma^{2}(R) = \int^{\infty}_{0}W^{2}(kR)\mathcal{P}_{\delta}(k)\frac{dk}{k},
\eeqa
where $W(kR)$ is the Fourier transform of the window function. $\mathcal{P}_{\delta}(k)$ is the power spectrum of 
the primordial density perturbations, and it is related to the power spectrum of primordial curvature perturbations, 
$\mathcal P_{\mathcal R}(k)$ as~\cite{Josan:2009} 

\beqa
\mathcal P_{\delta}(k) = \frac{16}{3}\left(\frac{k}{aH}\right)^2 j_{1}^2(
k/\sqrt{3}aH)\mathcal P_{\mathcal R}(k),
\eeqa
where $j_{1}$ is a spherical Bessel function. Different inflation models predict different forms of 
$\mathcal P_{\mathcal R}(k)$. For the general slow-roll inflation models, $\mathcal P_{\mathcal R}(k)$ can be written as~\cite{Kohri,Josan:2009,Leach}

\beqa
\mathcal P_{\mathcal R}(k) = \mathcal P_{\mathcal R}(k_{0})\left(\frac{k}{k_{0}}\right)^{n(k_{0})-1}.
\eeqa
We use this form for our calculations and for more detailed discussions one can refer to e.g. Refs.~\cite{Josan:2009,Bringmann_1,fangdali}. 

\begin{figure}
\epsfig{file=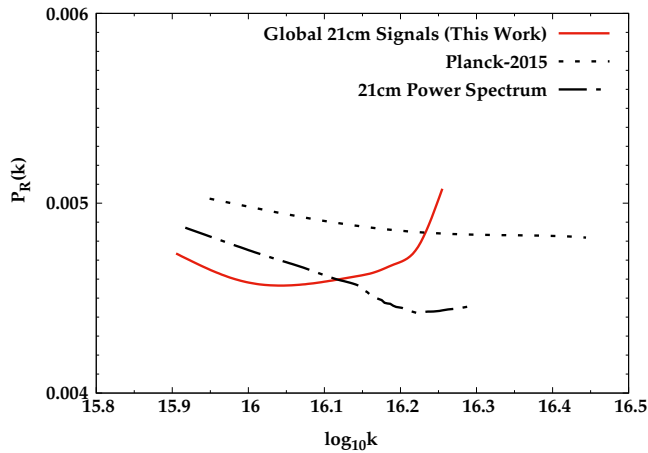,width=0.5\textwidth}
\caption{Constraints on the power spectrum of the primordial curvature perturbations, $\mathcal P_{\mathcal R}(k)$, 
by requiring the bright temperature of the global 21cm signals in the redshift range of $10\lesssim z\lesssim 30$ as $\delta T_{b}\lesssim -100 \rm mK$ (red solid line). 
For comparison, the constraints from the Placnk-2015 data~\cite{mnras} (black dotted line) 
and the potential 21cm power spectrum~\cite{mack_21} (black dashed-dotted line) are also shown.}
\label{fig:PPSC}
\end{figure}

Using the above equations and the constraints on the initial mass fraction of PBHs, in Fig.~\ref{fig:PPSC}, 
we display the upper limits on the power spectrum of primordial curvature perturbations for the scale range 
$8.0\times 10^{15} \lesssim k \lesssim 1.8\times 10^{16}$, corresponding to the mass range considered here. 
The strongest upper limit is 
$\mathcal P_{\mathcal R}(k) \sim 0.0046$. For comparison, in Fig.~\ref{fig:PPSC}, we also display the constraints 
from the Planck-2015 and 21cm power spectrum, corresponding to the constraints on $\beta_{\rm PBH}$ shown in Fig.~\ref{fig:dPBHSLCF}. 
Similar to the discussions about the constraints on $\beta_{\rm PBH}$ in the previous section, for the scale range considered here, since we have focused on 
the effects of PBHs on the IGM in the redshift range $10\lesssim z\lesssim 30$, 
the strongest upper limit appears at the smaller 
value of $k\sim 1.1\times 10^{16} ~\rm Mpc^{-1}$ with $\mathcal P_{\mathcal R}(k) \sim 0.0046$, 
corresponding to a larger PBH mass. For the potential constraints 
from the 21cm power spectrum, since the authors focused on the redshift range $30\lesssim z\lesssim 300$, 
the strongest upper limit appears at the larger value of $k\sim 1.7\times 10^{16} ~\rm Mpc^{-1}$ 
with $\mathcal P_{\mathcal R}(k) \sim 0.0044$, which corresponds to 
a smaller PBH mass. The constraints on $\mathcal P_{\mathcal R}(k)$ from the CMB are mainly from 
higher redshifts, 
and the strongest upper limit appears at the larger value of $k\sim 2.8\times 10^{16} ~\rm Mpc^{-1}$ 
with $\mathcal P_{\mathcal R}(k) \sim 0.0048$, corresponding to a smaller PBH mass.

\section{conclusions}

We have investigated the influence of PBHs on the IGM due to Hawking radiation for the mass range 
$6\times 10^{13} {\rm g} \lesssim M_{\rm PBH}\lesssim 3\times 10^{14} \rm g$. 
Particles emitted by PBHs interact with those that exist in the Universe. Due to these interactions, 
the degree of ionization and the temperature of the IGM are 
enhanced after a redshift $z\sim 1100$. The changes of the evolution of the IGM have an influence on the astrophysical 
observations such as the global 21cm 
signal. Inspired by the recent observations on the global 21cm signal in the redshift range $10\lesssim z\lesssim 30$ 
by EDGES, we have investigated the effects of PBHs on the global 21cm signal. We have found that the main effect is 
to suppress the absorption amplitude of the global 21cm signal, which is consistent with previous works. By requiring 
that the differential brightness temperature of the global 21cm signal be $\delta T_{b}\lesssim -100~\rm mK$, 
we obtained the upper limits on the 
initial mass fraction of PBHs depending on their mass. The strongest upper limit is $\beta_{\rm PBH} \sim 2\times 10^{-30}$. 
In previous works, for the same mass range of PBHs, the constraints on $\beta_{\rm PBH}$ can also be obtained 
using the CMB data, the extragalactic photon background and the potential 21cm power spectrum. By comparing these 
constraints, we found that for the mass range of PBHs considered here the global 21cm signal or the 21cm power spectrum 
could give the strongest upper limit. Since the formation of PBHs is related to the primordial 
curvature perturbations, using the constraints 
on the initial mass fraction of PBHs we obtained the upper limits on the power spectrum of primordial curvature perturbations for 
the scale range $8.0\times 10^{15} \lesssim k \lesssim 1.8\times 10^{16}~\rm Mpc^{-1}$, corresponding to 
the mass range of PBHs considered here. The strongest upper limit is $\mathcal P_{\mathcal R}(k) \sim 0.0046$. 

Previous works (e.g.,~Ref.~\cite{yinzhema}) also investigated the limits on PBHs by requiring 
$\delta T_{b} \lesssim -50~\rm mK$. The constraints on the abundance of PBHs are about a factor of 3 weaker for $\delta T_{b} 
\lesssim -50~\rm mK$ than that of $\delta T_{b}\lesssim -100~\rm mK$. These differences should have a slight influences on the 
constraints of primordial curvature perturbations. In this work, inspired by the observational results of EDGES, 
we focused on the global 21cm signal 
in the redshift range $10\lesssim z\lesssim 30$. As shown in Fig.~\ref{fig:delta_eg}, there are also global 21cm signals 
in the redshift range $30\lesssim z\lesssim 300$, and observing these global 21cm signals is very difficult. 
Future experiments that could be run, e.g., on the Moon would detect these global 21cm signals~\cite{moon_21}. As discussed 
in the above sections and motivated by the work of Ref.~\cite{mack_21}, the expected 21cm power spectrum in the redshift range $10\lesssim z\lesssim 30$ should give stronger upper 
limits on $\beta_{\rm PBH}$ or $\mathcal P_{\mathcal R}(k)$ than that of global 21cm signal. For the redshift range 
considered here, the global 21cm signal can also be influenced significantly by other astrophysical factors, such as 
the star formation efficiency, the collapse fraction of the halos and so on~\cite{cohen-mnras,Pritchard:2011xb,Furlanetto:2006jb}. 
We will investigate these issues in future work.

In summary, we investigated the influence of PBHs on the global 21cm signal at the redshift $z\sim 17$ 
due to Hawking radiation. Compared with previous works, the new features of this work are as follows:

\begin{itemize}
\item We have extended the mass range of PBHs to $6\times 10^{13} {\rm g} \lesssim M_{\rm PBH}\lesssim 3\times 10^{14} \rm g$. 
PBHs with masses in this range evaporate in the redshift range $10\lesssim z\lesssim 30$, and they are expected to have 
significant effects on the global 21cm signal depending on their initial mass fraction $\beta_{\rm PBH}$. 
Inspired by the observational results of the EDGES experiment, by requiring 
the differential brightness temperature of the global 21cm signal to be $\delta T_{b}\lesssim -100~\rm mK$, we have found that the strongest upper limit 
of the initial mass fraction of PBHs 
is $\beta_{\rm PBH} \sim 2\times 10^{-30}$, and as far as we know this is currently the 
strongest upper limit for the mass range considered here
\footnote{As shown in Fig.~\ref{fig:dPBHSLCF}, the upper limits from the future expected observations of the power spectrum of 
the 21cm signals could be stronger.}.  
 
\item Based on the constraints on the initial mass faction of PBHs, we obtained the upper limits 
on the power spectrum of primordial curvature perturbations for 
the scale range $8.0\times 10^{15} \lesssim k \lesssim 1.8\times 10^{16}~\rm Mpc^{-1}$. 
The strongest upper limit is $\mathcal P_{\mathcal R}(k) \sim 0.0046$, and this (as far as we know) 
is currently the best upper limit.

\item For the mass range considered in this work, and after comparing with other works, we found that the 
observations and studies on the global 21cm signal 
or the power spectrum of the 21cm signal could give the strongest upper limits on the initial mass fraction of PBHs 
and the power spectrum of primordial curvature perturbations. Moreover, since the astrophysical influence 
on the 21cm signals are very weak in the redshift range $30\lesssim z\lesssim 300$, future observations of the 21cm signals in this redshift range (e.g., experiment on the Moon) are very useful for the researches on PBHs.

\end{itemize}

\section{Acknowledgements}
This work is supported in part by the National Natural Science Foundation of China 
(under Grants No.11505005 and No.U1404114). Y. Yang is supported in part by the Youth Innovations and Talents Project of Shandong 
Provincial Colleges and Universities (Grant No. 201909118)
\

\bibliographystyle{apsrev4-1}

%

\end{document}